\documentclass[twocolumn,english,aps]{revtex4}
\usepackage[T1]{fontenc}
\usepackage[latin9]{inputenc}
\usepackage{textcomp}
\usepackage{amsmath}
\usepackage{color}
\usepackage{graphicx}
\usepackage{amssymb}
\usepackage{esint}

\makeatletter
\@ifundefined{textcolor}{}
{%
 \definecolor{BLACK}{gray}{0}
 \definecolor{WHITE}{gray}{1}
 \definecolor{RED}{rgb}{1,0,0}
 \definecolor{GREEN}{rgb}{0,1,0}
 \definecolor{BLUE}{rgb}{0,0,1}
 \definecolor{CYAN}{cmyk}{1,0,0,0}
 \definecolor{MAGENTA}{cmyk}{0,1,0,0}
 \definecolor{YELLOW}{cmyk}{0,0,1,0}
 }

\makeatother

\makeatother

\usepackage{babel}

\begin{document}

\preprint{This line only printed with preprint option}

\title{Electrical detection of free induction decay and Hahn echoes in phosphorus
doped silicon}

\author{Jinming Lu}

\affiliation{Walter Schottky Institut, Technische Universit\"{a}t M\"{u}nchen, Am Coulombwall
3, 85748 Garching, Germany}

\author{Felix Hoehne}

\email[corresponding author, email: ]{hoehne@wsi.tum.de}

\affiliation{Walter Schottky Institut, Technische Universit\"{a}t M\"{u}nchen, Am Coulombwall
3, 85748 Garching, Germany}

\author{Andre R. Stegner}

\affiliation{Walter Schottky Institut, Technische Universit\"{a}t M\"{u}nchen, Am Coulombwall
3, 85748 Garching, Germany}

\author{Lukas Dreher}

\affiliation{Walter Schottky Institut, Technische Universit\"{a}t M\"{u}nchen, Am Coulombwall
3, 85748 Garching, Germany}

\author{Hans Huebl}

\affiliation{Walther-Meissner-Institut, Bayerische Akademie der Wissenschaften, Walther-Meissner-Strasse 8, 85748 Garching, Germany}

\author{Martin Stutzmann}

\affiliation{Walter Schottky Institut, Technische Universit\"{a}t M\"{u}nchen, Am Coulombwall
3, 85748 Garching, Germany}

\author{Martin S.~Brandt}

\affiliation{Walter Schottky Institut, Technische Universit\"{a}t M\"{u}nchen, Am Coulombwall
3, 85748 Garching, Germany}
\begin{abstract}
Paramagnetic centers in a solid-state environment usually give rise
to inhomogenously broadened electron paramagnetic resonance (EPR)
lines, making conventionally detected free induction decay (FID) signals
disappear within the spectrometer dead time. Here, experimental results of an electrically detected FID
of phosphorus donors in silicon epilayers with natural isotope composition
are presented, showing Ramsey fringes within the first $150$\,ns.
An analytical model is developed to account for the data obtained
as well as for the results of analogous two-pulse echo experiments. The
results of a numerical calculation are further presented to assess
the capability of the method to study spin-spin interactions. 
\end{abstract}
\maketitle

\section{introduction}

Solid-state based quantum computer (QC) architectures have recently
attracted considerable interest due to the prospect of possible scalability.
Several implementations have been proposed, such as superconducting
tunnel junctions~\cite{Makhlin01SuperconductingQC}, nuclear spins
in crystal lattices~\cite{Yamaguchi1999LatticQC}, nuclear spins
of donors in Si~\cite{Kane98}, and electron spins localized in quantum
dots (QDs)~\cite{Loss98QC,Imamoglu1999QDS} or at donors~\cite{Vrijen2000SiGeQC}.
However, a realistic solid-state environment introduces an intrinsic
inhomogeneity of the qubit properties due to impurities, defects,
interfaces etc., resulting in possible computational errors. Nevertheless,
fault-tolerant error correction schemes promise to compensate such
errors up to a certain level~\cite{DiVincenzo1996ErrorCorr,Knill1998ErrorCorr,Steane1999ErrorCorr},
making a quantification of the inhomogeneity a crucial step in the
assessment of the suitability of specific physical realizations of
qubits for QC.

In the case of QDs and donors, the Zeeman levels of the electron spin
in an external magnetic field are used as the qubit states. The Zeeman
energy of each electron spin will vary within a spin ensemble, e.g.~due to the random orientation of nuclear spins in the lattice that
generate a spatially fluctuating local magnetic field at the position
of each electron spin. The resulting distribution $\Phi(\omega_{\mathrm{S}})$
of the Larmor frequencies $\omega_{\mathrm{S}}$ of the electron spins
can be quantified in the frequency domain by the inhomogenous linewidth
obtained from a traditional continuous wave (cw) electron paramagnetic
resonance (EPR) experiment\,\cite{Abe2010}. Alternatively,
a time domain analysis of $\Phi(\omega_{\mathrm{S}})$ can be carried
out by measuring the decay of the transverse magnetization in the
free induction decay (FID) of a conventional pulsed EPR experiment.
The problem encountered in both methods is their respective detection
limit which is far above the number of spins present in nanoscale
structures proposed for QCs. Moreover, FID has not seen a widespread
application for studies of paramagnetic species in solids since the
broad Larmor frequency distribution usually encountered in solid-state
spin systems makes the FID signal disappear on a time scale that is
usually much shorter than the spectrometer dead time $t_{\mathrm{D}}$
after a microwave pulse of typically $100-200$\,ns \cite{Schweiger01,Dikanov92ESEEM,Blok2009FIDhighMW}.

Optically and electrically detected magnetic resonance experiments
(ODMR and EDMR, respectively) have shown much promise since they exceed
conventional EPR concerning detection sensitivity by at least six
orders of magnitude~\cite{Jelezko04SingleSpin,mccamey06fewspins}.
In the case of phosphorus donors ($^{31}$P) in silicon, the development
of pulsed EDMR (pEDMR)~\cite{Boehme01EDMR,Boehme03EDMR,Stegner06}
has paved the way for studying dynamical properties of spin systems
by applying different microwave pulse sequences, such as using the
Hahn-echo sequence to study coherence times $T_{2}$~\cite{Huebl08Echo},
or the inversion recovery sequence to study spin-lattice relaxation
times $T_{1}$~\cite{Paik2010T1T2}. The underlying principle of
pEDMR on $^{31}$P donors is the use of microwave pulses for the generation
of coherent states of $^{31}$P spins, which can be detected by transient
photoconductivity measurements~\cite{Boehme03EDMR,Stegner06}, making
use of spin-dependent recombination of spin pairs formed by $^{31}$P
and paramagnetic Si/SiO$_{2}$ interface states~\cite{Kaplan78Spindep,Hoehne09KSM}.

In current pulsed EPR, the dead-time problem of FID has been circumvented
by several techniques such as high-frequency EPR~\cite{Blok2009FIDhighMW},
FID-detected hole burning~\cite{Schweiger01} or detection of the
EPR spectra by the Electron Spin Echo (ESE), although the latter is
more sensitive to distortions through nuclear modulations in contrast
to FID~\cite{Schweiger01,Blok2009FIDhighMW}. In this paper, we study
the possibility of electrically detecting (ED) FID using a tomography
technique developed earlier \cite{Huebl08Echo} to investigate the
Larmor frequency distribution $\Phi(\omega_{\mathrm{S}})$ of a $^{31}$P
ensemble in natural silicon ($^{\mathrm{nat}}$Si). It is shown that
information within the usual dead time of conventional EPR-detected
FID can be obtained. An analytical equation is deduced to describe
the experimental data, which in turn agrees well with the results
of continuous wave (cw) EDMR experiments. Furthermore, a numerical
study is performed to assess the EDFID technique in terms of its capability
to quantitatively investigate the coupling of spin pairs.

\section{experimental details}
\label{ss:ExpDetails}

The sample used in this work is fabricated by chemical vapor deposition
and consists of a $22$\,nm thick $^{\mathrm{nat}}$Si layer with
$\mathrm{[P]}=9\times10^{16}$\,cm$^{-3}$ covered by a native
oxide. It is grown on a $2.5$\,\textmu{}m thick, nominally undoped
$^{\mathrm{nat}}$Si buffer on a silicon-on-insulator substrate. Evaporated
interdigit Cr/Au contacts with a period of $20$\,\textmu{}m
covering an active area of $2\times2.25$\,mm$^{2}$ are biased with
$100$\,mV, resulting in a current of $\approx80$\,\textmu{}A under illumination
with white light. The major paramagnetic states are identified in
cwEDMR experiments performed in a Bruker X-band dielectric microwave resonator for pulsed EPR. The microwave frequency of
$\nu_{\mathrm{mw}}=\omega_{\mathrm{mw}}/(2\pi)=9.7400$\,GHz is provided
by an HP83640 microwave synthesizer. The measurements are performed
at $5.5$\,K in a helium gas flow cryostat. The samples are oriented in an external magnetic field $B_0$ with the [001] axis of the Si wafer parallel to $B_0$. 
\begin{figure}[h]
\begin{centering}
\includegraphics[width=8cm]{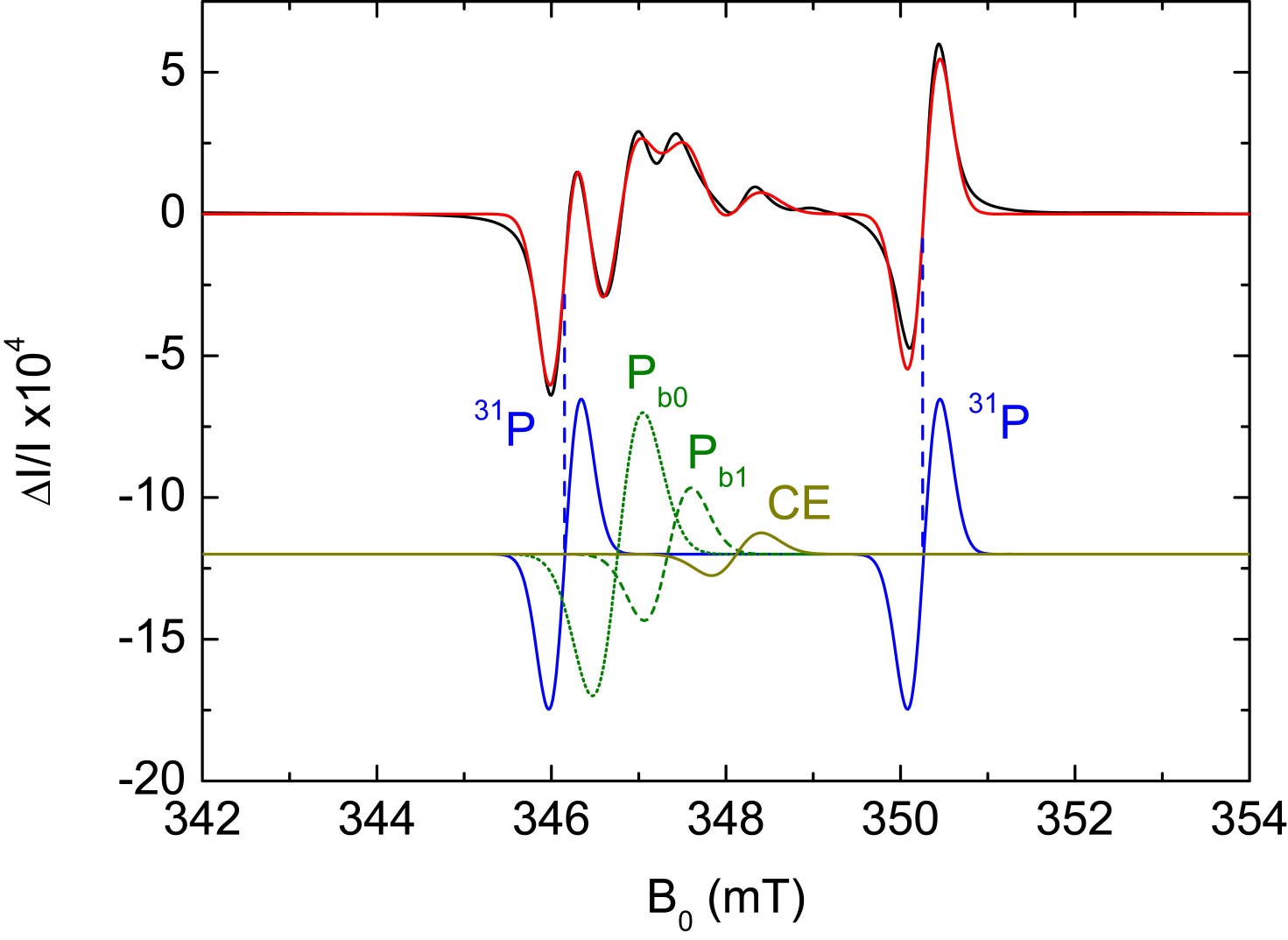}
\par\end{centering}
\caption{\label{fig:cwSpektrum}First derivative spectrum of the relative change
$\Delta I/I$ of the photocurrent in a cwEDMR experiment on $^{\mathrm{nat}}$Si:P
with a magnetic field modulation amplitude of $0.2$\,mT. The red curve
represents a fit using Gaussian lineshapes and equal amplitudes of both $^{31}$P resonance lines.
The constituent lines of the fit are shown below for better visibility.}

\end{figure}
The first derivative of the relative current change $\Delta I/I$ {[}cf. Fig.\,\ref{fig:cwSpektrum}{]}
is measured with magnetic field modulation and lock-in detection as
a function of $B_{0}$ provided by a Bruker
BE 25 electromagnet. We used the two pronounced $^{31}$P hyperfine resonance lines (denoted hf($^{31}$P)) at $B_{0}=346.17$\,mT
and $B_{0}=350.27$\,mT to calibrate the magnetic field $B_0$ such that their center field corresponds to $g=1.9985$\,\cite{Feher59I}.
In addition, a resonance line at $g=2.0069\pm0.0004$ arising from the P$_{\mathrm{b0}}$ center
at the Si/SiO$_{2}$ interface can be observed in accordance with
previous studies for $B_0$||[001]\,\cite{poindexter81db}. A further resonance line is observed at $g=2.0036\pm0.0004$ which, due to its $g$-factor, is attributed to the P$_{\mathrm{b1}}$ center\,\cite{Stesmans98dbHF}. We are aware of the fact that there is no concensus in the literature whether the P$_{\mathrm{b1}}$ defect is electrically active\,\cite{Stesmans98dbDensity, Mishima99} and therefore, measurements of the angular dependence of the $g$-factor or the hyperfine interactions with $^{29}$Si nuclei would be needed for an unambiguous identification. The small resonance line at
$B_{0}=348.12$\,mT could be attributed to conduction band electrons (CE)
with a $g$-factor of $g\approx1.9990\pm0.0004$\,\cite{Young97gfactor}. A fit assuming Gaussian line shapes and equal amplitudes of both $^{31}$P resonance lines is indicated by the red line in Fig.\,\ref{fig:cwSpektrum}.
The decomposition of the fitted spectrum is shown in the lower
part of Fig.\,\ref{fig:cwSpektrum}. 
An analysis of the isolated high-field hf($^{31}$P) line yields a
peak-to-peak line width of $\Delta B_{\mathrm{PP}}^{\mathrm{cwEDMR}}=0.30\pm0.03$\,mT after
correcting for the influence of magnetic field modulation \cite{Poole67ESR}.
This line broadening is predominantly inhomogeneous, caused by randomly
oriented nuclear spins of the $^{29}$Si isotope in natural Si, which
give rise to an unresolved superhyperfine multiplet~\cite{Abe05Linewidth}.
Homogeneous broadening only plays a minor role since $T_{1}$ and $T_{2}$
times of the $^{31}$P donor spins determined by ED inversion recovery~\cite{Paik2010T1T2} and ED Hahn echo~\cite{Huebl08Echo} experiments
on this sample yield $T_{1}\approx5.3$\,\textmu{}s and $T_{2}\approx3.3$\,\textmu{}s,
respectively, when the measured decays are fitted by a single exponential dependence.

The pulsed EDMR experiments are performed at a microwave frequency
of $\nu_{\mathrm{mw}}=\omega_{\mathrm{mw}}/(2\pi)=9.7331$\,GHz under the same orientation of the sample as the cwEDMR experiments. For all pulsed measurements, the magnetic field $B_0$ is corrected by the magnetic field offset determined by cwEDMR. The
microwave pulses are shaped using a SPINCORE PulseBlasterESR-Pro 400
MHz pulse generator and a system of microwave mixers, and are then
amplified by an Applied Systems Engineering 117X traveling wave tube
with a maximum peak power of 1\,kW. The actual pulse shapes coupled
into the resonator are checked in reflection experiments. The quality
factor of the dielectric resonator is adjusted to make a compromise
between sufficient excitation bandwidth and a microwave magnetic field
$B_{1}$ high enough for coherent spin manipulation. The
adjustment of the microwave power is achieved by a tuneable attenuator;
the $\pi/2$-pulse time of $\tau_{\pi/2}=15$\,ns corresponding to
$B_{1}=0.6$\,mT used throughout this paper is determined
in Rabi-oscillation experiments as previously developed~\cite{Boehme03EDMR,Stegner06}.
The current transients are acquired using a current amplifier followed
by a voltage amplifier and a fast data aquisition card. To improve
the signal-to-noise ratio, we applied a two-step phase cycling sequence
where the phase of the last $\pi/2$ pulse was switched by 180$^{\circ}$ at a frequency of $\approx5$\,Hz and the signals for each
phase are then subtracted from each other. Switching the phases at
a frequency of several Hz has the same purpose as a lock-in detection scheme
and improves the signal-to-noise ratio by reducing the effects of
$1/f$-noise in the measurement setup.

\section{results and discussion}


\subsection{Electrically detected FID}

\begin{figure}[h]
 \begin{centering}
\includegraphics[width=8cm]{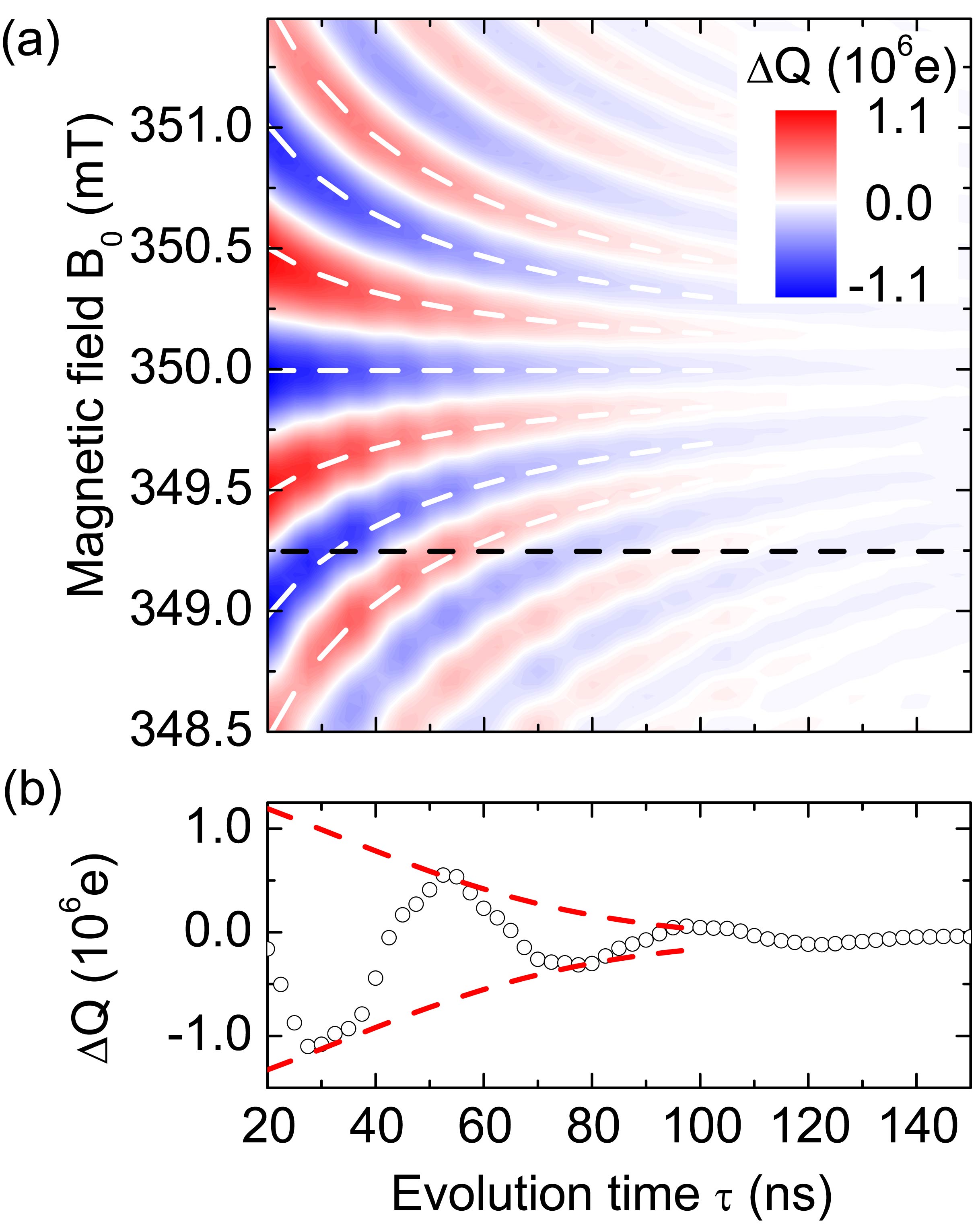} 
\par\end{centering}
\caption{\label{fig:RamseyResult}
Electrically detected free induction decay
or Ramsey experiment on the high-field hf($^{31}$P) resonance. (a)
Contour plot of $\Delta Q$ as a function of the external magnetic
field $B_{0}$ and the free evolution time $\tau$. White dashed curves
mark the positions of local extrema of $\Delta Q$ described by Eq.\,(\ref{eq:hyperbolas}).
The black dashed line indicates the position where the cross-sectional
chart shown in panel (b) is taken. (b) Cross-section of $\Delta Q$
along the evolution time axis at $B_{0}=349.25$ mT. Red dashed curves
illustrate the Gaussian-shape damping of the oscillation amplitude.}
\end{figure}

The EDFID tomography is performed by a $\pi/2$-$\tau$-$\pi/2$ pulse
sequence with varying evolution time $\tau$, consisting of the conventional
free induction pulse sequence $\pi/2\mbox{-}\tau$ followed by a $\pi/2$-projection
pulse as usually applied in multi-pulse EDMR experiments \cite{Huebl08Echo,Paik2010T1T2}.
Hence, it coincides with the pulse sequence of the Ramsey experiment
\cite{Ramsey50}. The photocurrent transients measured for each $B_{0}$
and $\tau$ are integrated over a fixed recording time interval satisfying
the criteria described in Ref.\,\cite{Gliesche2008EffectCoupling},
resulting in a charge difference $\Delta Q$. Figure\,\ref{fig:RamseyResult}
shows experimental results of an EDFID tomography experiment on the
isolated high-field hf($^{31}$P) line. 

We will now show that the pattern in Fig.\,\ref{fig:RamseyResult}\,(a),
which is characteristic of an EDFID can be understood by a simple
model in which the contribution of the state of each spin pair at
the end of the second $\pi/2$ pulse is proportional to its projection
onto the singlet state $|S\rangle$ \cite{Boehme03EDMR,Stegner06}.
Hence, the measured charge $Q\propto-S^{\mathrm{av}}(\tau)=-\mathrm{Tr}(|S\rangle\langle S|\hat{\rho})$
reveals the average singlet content of the spin pair ensemble described
by the density operator $\hat{\rho}$. This is in contrast to conventional
ESR, where for an FID the magnetization after a $\pi/2$ pulse is
detected. For microwave frequencies close to the Larmor frequency
of the high-field hf($^{31}$P), the singlet content $S(\tau)$ of
each spin pair reflects the dynamics of only this spin species \cite{Boehme03EDMR}
while in a first approximation the spin state of P$_{\mathrm{b0}}$
is unaltered and just serves as a projection partner. This is justified
since the separation of the Larmor frequencies of the $^{31}$P and
P$_{\mathrm{b0}}$ spins for the high-field hf($^{31}$P) resonance is approximately
one order of magnitude larger than the on-resonance Rabi frequency
$\omega_{1}=g\mu_{\mathrm{B}}B_{1}/\hbar$. The minor effects of the off-resonance excitation of the other
resonance lines can be seen as small oscillations on the Ramsey pattern in Fig.\,\ref{fig:RamseyResult}\,(a)
at magnetic fields lower than 350\,mT. 
We also neglect spin-spin interaction
and incoherent processes during the pulse sequence. The former will
be addressed in Sec.~\ref{ss:Spin-spin coupling} and the latter
is a valid assumption since the time constant for the fastest incoherent
process is $T_{2}\approx3.3$\,\textmu{}s as measured in ED Hahn
echo decay experiments on this sample. With these assumptions, an
expression for the theoretically expected signal \begin{equation}
\Delta Q\propto-\exp\left[-\frac{1}{2}\frac{\sigma_{\omega}^{2}\tilde{\omega}_{1}^{2}}{\sigma_{\omega}^{2}+\tilde{\omega}_{1}^{2}}\tau^{2}\right]\cos\left[\frac{\tilde{\omega}_{1}^{2}}{\sigma_{\omega}^{2}+\tilde{\omega}_{1}^{2}}\Delta\omega\tau\right]\label{eq:ramseysig}\end{equation}
with $\tilde{\omega}_{1}=\omega_{1}/\sqrt{2}$ and $\Delta\omega=\omega_{0}-\omega_{\mathrm{mw}}$
can be derived following Ref.\,\cite{Jaynes55matrix, Bloom55Echo} as shown in Appendix\,\ref{sec:App}. In Eq.\,(\ref{eq:ramseysig}),
$\sigma_{\omega}$ quantifies the width of the Larmor frequency distribution
as defined in Eq.\,(\ref{eq:gaussian}). The locations of the local
extrema of $\Delta Q$ are given by Eq.\,(\ref{eq:hyperbolas2}) as
\begin{equation}
B_{0}-B_{\mathrm{res}}=\frac{n\pi\hbar\left(1+2(\sigma_{\omega}/\omega_{1})^{2}\right)}{g\mu_{\mathrm{B}}}\frac{1}{\tau},\qquad n\in\mathbb{Z},\label{eq:hyperbolas}\end{equation}
representing hyperbolas in the $B_{0}$-$\tau$-plane. These hyperbolas
fit the experimentally observed pattern well, as evident from the
white dashed curves in Fig.\,\ref{fig:RamseyResult}~(a). The exponential
term in Eq.\,(\ref{eq:ramseysig}) describes an envelope in the time
domain which shows an $\exp\left[-(\tau/T_{\mathrm{FID}})^{2}\right]$-type
decay behavior \cite{DeVoe78FID} as depicted by red dashed line in Fig.\,\ref{fig:RamseyResult}~(b) with the time constant \begin{equation}
T_{\mathrm{FID}}=\sqrt{\frac{2}{\sigma_{\omega}^{2}}+\frac{4}{\omega_{1}^{2}}}\,.\label{eq:Ramseytimeconst}\end{equation}
This implies that for short pulses, i.e.~in the high microwave power limit,
$T_{\mathrm{FID}}$ is inversely proportional to the width of the
Larmor frequency distribution and thus $\Delta B_{\mathrm{PP}}^{\mathrm{cwEDMR}}$
of the hf($^{31}$P) resonance line in cwEDMR. The actual decay characteristics
deviate from the $\exp\left[-(\tau/T_{\mathrm{FID}})^{2}\right]$ behavior since
the lineshape is a convolution of a Gaussian and Lorentzian lineshape rather than a pure Gaussian.%
\begin{figure}[h]

\begin{centering}
\includegraphics[width=8cm]{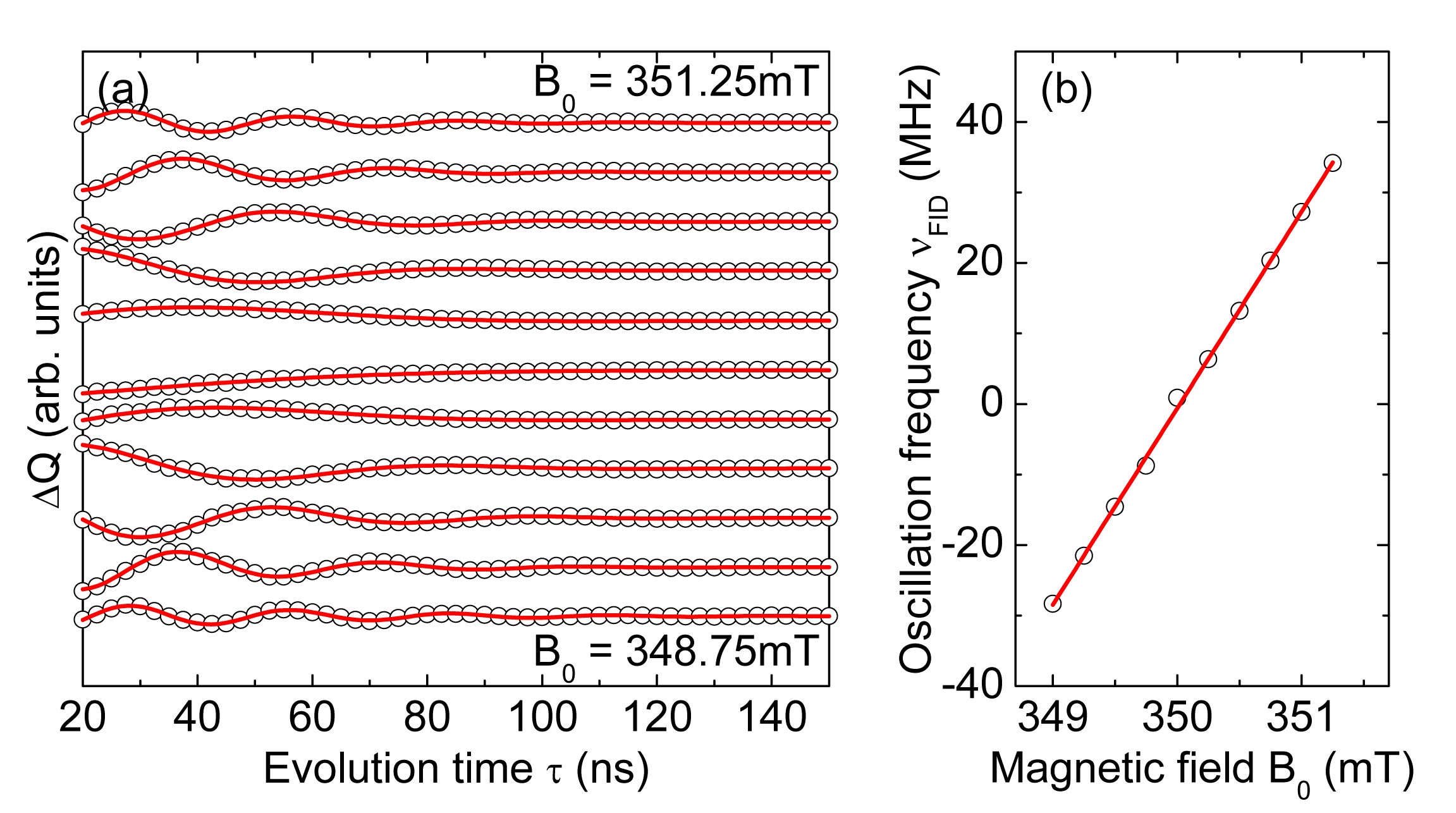} 
\par\end{centering}

\centering{}\caption{\label{fig:Ramseyfreq}Oscillations in the evolution time domain.
(a) Cross-sections of $\Delta Q$ along the evolution time axis taken
at different values of $B_{0}$. The damping of the oscillations indicates
that the spin ensemble dephases. Red curves are fits based on the
model given by Eq.\,(\ref{eq:Ramseyosctime}). (b) Frequency $\nu_{\mathrm{FID}}$ of the
damped oscillations as a function of the external magnetic field $B_{0}$
shows a linear behavior as expected from Eq.\,(\ref{eq:nuramseyMF}).}

\end{figure}

In Fig.\,\ref{fig:Ramseyfreq}\,(a), cross-sections of $\Delta Q$
along the evolution time axis taken at different values of $B_{0}$
are plotted as a function of the evolution time $\tau$, revealing
strongly damped oscillations.
These characteristics are consistent with those described by Eq.\,(\ref{eq:ramseysig}).
A clear dependence of the frequencies of the damped oscillations $\nu_{\mathrm{FID}}$ on the
external magnetic field $B_{0}$ can be observed. The red lines in
Fig.\,\ref{fig:Ramseyfreq}(a) show fits of the oscillations by the
function
\begin{equation}
\Delta Q\sim-Ae^{-\left(\frac{\tau+\tau_{0}}{T_{\mathrm{FID}}}\right)^{2}}\cos\left[2\pi\nu_{\mathrm{FID}}(\tau+\tau_{0})\right]
\label{eq:Ramseyosctime}
\end{equation}
which is based on the model given in Eq.\,(\ref{eq:ramseysig}).
In all the fits shown in Fig.\,\ref{fig:Ramseyfreq}\,(a), a global
phase correction of $\tau_{0}\approx20$\,ns has to be taken into
account which is comparable to the overall pulse length of 30~ns.
It can be attributed to the fact that dephasing during the pulse times
can not be neglected due to the finite pulse width compared to the
evolution time $\tau$. The values of $\nu_{\mathrm{FID}}$ obtained
from the fits are plotted as a function of the external magnetic field
$B_{0}$ and displayed in Fig.\,\ref{fig:Ramseyfreq}\,(b). A clear
linear dependence of $\nu_{\mathrm{FID}}$ on the magnetic field $B_{0}$
can be observed as expected from Eq.\,(\ref{eq:ramseysig}), 
\begin{equation}
\nu_{\mathrm{FID}}=\frac{1}{2\pi}\frac{\tilde{\omega}_{1}^{2}}{\sigma_{\omega}^{2}+\tilde{\omega}_{1}^{2}}\left(\omega_{0}-\omega_{\mathrm{mw}}\right)=k\left(B_{0}-B_{\mathrm{res}}\right),
\label{eq:nuramseyMF}
\end{equation}
with $k=g\mu_{\mathrm{B}}/[h(1+2(\sigma_{\omega}/\omega_{1})^{2})]$.
This is consistent with the linear dependence of the oscillation frequency
on the detuning in a Ramsey experiment \cite{Plourde2005Ramsey,Wallraff05SCRamsey}.
From a linear fit of the data the value of $B_{\mathrm{res}}=350.02\pm0.01$
mT is obtained, which corresponds to the center position of the high-field
hf($^{31}$P) resonance line. Using Eq.\,(\ref{eq:Ramseytimeconst})
and $2\pi/\omega_{1}=60$\,ns, the average value  $T_{\mathrm{FID}}=64.4\pm5$\,ns
obtained from the fits of the damped oscillations can be related to
an expected cwEDMR peak-to-peak line width of  $\Delta B_{\mathrm{PP}}^{\mathrm{FID}}=2\hbar\sigma_{\omega}/(g\mu_{\mathrm{B}})=0.26\pm0.02\mbox{ mT}$.
This is in agreement with the result obtained from cwEDMR $\Delta B_{\mathrm{PP}}^{\mathrm{cwEDMR}}=0.30\pm0.03\mbox{ mT}$,
demonstrating the consistency of the experiments.

\subsection{Electrically detected Hahn echo.jpg}

Electrically detected echo sequences have been previously used to
study $T_{2}$~\cite{Huebl08Echo} and $T_{1}$ times \cite{Paik2010T1T2}
of phosphorus donors near Si/SiO$_{2}$ interface defects. In this
section, we will present detailed experimental results of the electrically
detected Hahn echo measurement with a focus on the fine structure
of the echo response.

The echo is measured using the previously developed tomography technique
\cite{Huebl08Echo} extending the pulse sequence $\pi/2$-$\tau_{1}$-$\pi$-$\tau_{2}$
of the conventional two-pulse spin echo containing two free evolution
times $\tau_{1}$ and $\tau_{2}$ by a final $\pi/2$ pulse as applied
in the EDFID technique. The measurements are conducted under the same
experimental conditions as the EDFID experiments. %
\begin{figure}[h]
\begin{centering}
\includegraphics[width=8cm]{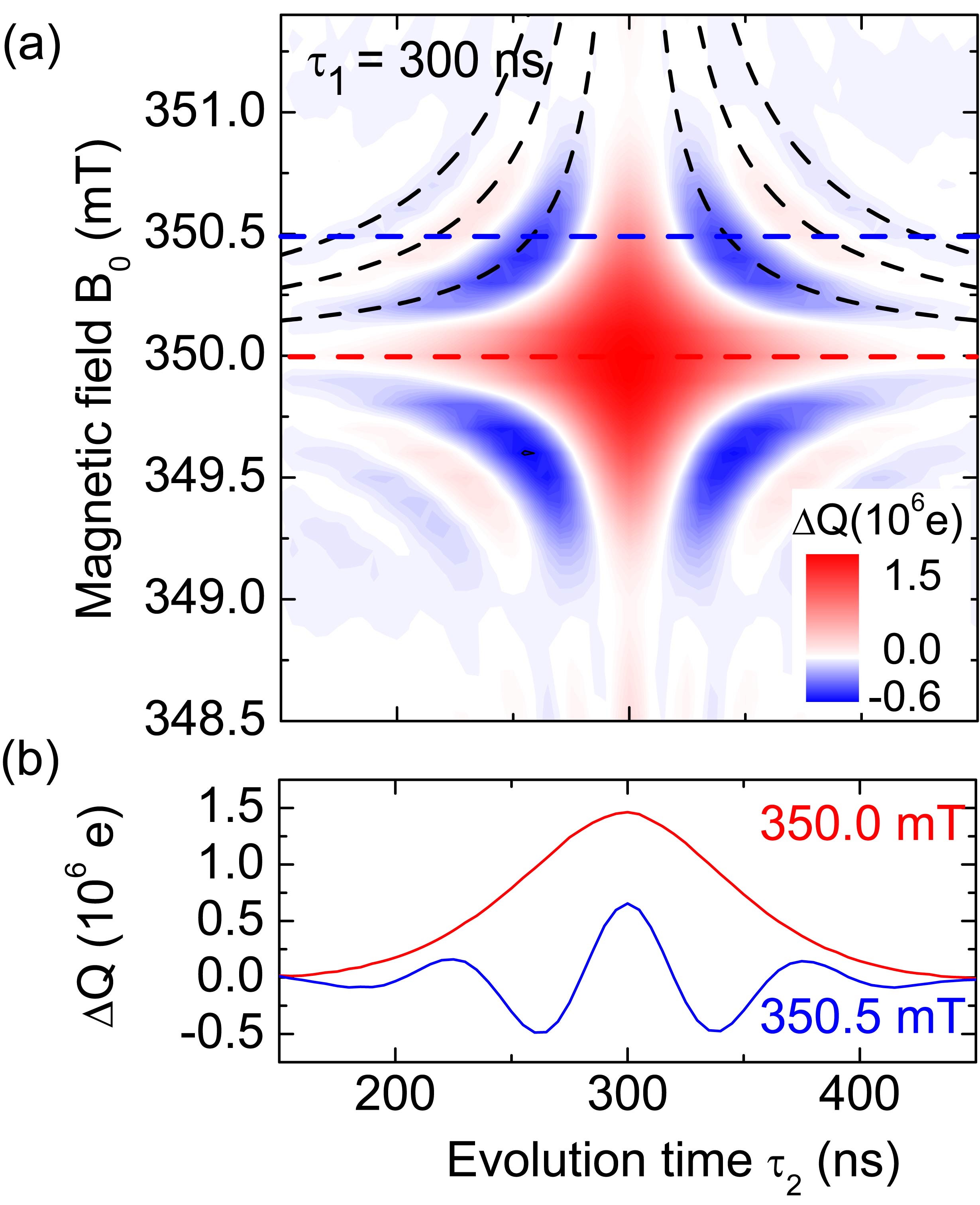} 
\par\end{centering}
\caption{\label{fig:Echo}
Electrically detected two-pulse Hahn echo on the
high-field hf($^{31}$P) resonance with $\tau_{1}=300$\,ns. (a)
Contour plot of $\Delta Q$ as a function of the external magnetic
field $B_{0}$ and the free evolution time $\tau_{2}$. Black dashed
lines mark the hyperbola pattern according to Eq.\,(\ref{eq:DeltaQEcho}).
Red and blue dashed lines indicate positions where cross-sectional
diagrams shown in panel (b) are taken. (b) Cross-section of $\Delta Q$
along the evolution time axis at resonance condition (red line, $B_{0}=350.0$\,mT)
and for the off resonant case (blue line, $B_{0}=350.5$\,mT). Please
refer to the text for details.}
\end{figure}

Figure\,\ref{fig:Echo} shows experimental results of an ED spin
echo tomography experiment on the isolated high-field hf($^{31}$P)
line with $\tau_{1}=300$\,ns held fix. The values of $\Delta Q$
plotted as a function of $B_{0}$ and $\tau_{2}$ {[}Fig.\,\ref{fig:Echo}(a){]}
are obtained in the same way as described in the EDFID section. Cross-sections along the evolution time axis are displayed
in Fig.\,\ref{fig:Echo}(b). The red curve taken at the resonance
field $B_{0}=350.0$\,mT shows a Gaussian-shaped peak centered around
$\tau_{2}=\tau_{1}=300$\,ns. The cross section of $\Delta Q$ at
the off-resonance field $B_{0}=350.5$\,mT, which is represented
by the blue curve, shows oscillations as a function of the evolution
time $\tau_{2}$ with a maximum at $\tau_{2}=\tau_{1}=300$\,ns,
which decay for $|\tau_{2}-\tau_{1}|\gg100$\,ns.

The characteristic pattern indicated by the black dashed hyperbolas
can be understood by the same quantitative model described in the previous
section on EDFID. The singlet content $S(\tau)$ proportional to the
recombination probability $P_{\uparrow,\downarrow}$ of a single spin
after a $\pi/2$-$\tau_{1}$-$\pi$-$\tau_{2}$-$\pi/2$ pulse sequence
can be calculated using the matrix formalism described in Ref.~\cite{Jaynes55matrix,Bloom55Echo}.
For a Larmor frequency distribution modelled by a Gaussian with standard
deviation $\sigma_{\omega}$ centered about $\omega_{0}$, we can
derive analogously to Eq.\,(\ref{eq:ramseysig2}) that \begin{eqnarray}
\Delta Q & \propto & \exp\left[-\frac{1}{2}\frac{\sigma_{\omega}^{2}\bar{\omega}_{1}^{2}}{\sigma_{\omega}^{2}+\bar{\omega}_{1}^{2}}(\tau_{2}-\tau_{1})^{2}\right]\nonumber \\
 &  & \times\cos\left[\frac{\bar{\omega}_{1}^{2}}{\sigma_{\omega}^{2}+\bar{\omega}_{1}^{2}}(\omega_{0}-\omega_{\mathrm{mw}})(\tau_{2}-\tau_{1})\right]\label{eq:DeltaQEcho}\end{eqnarray}
 with $\bar{\omega}_{1}=\omega_{1}/2$ ~\cite{factorsqrt2}. 
\begin{figure}[h]
\begin{centering}
\includegraphics[width=8cm]{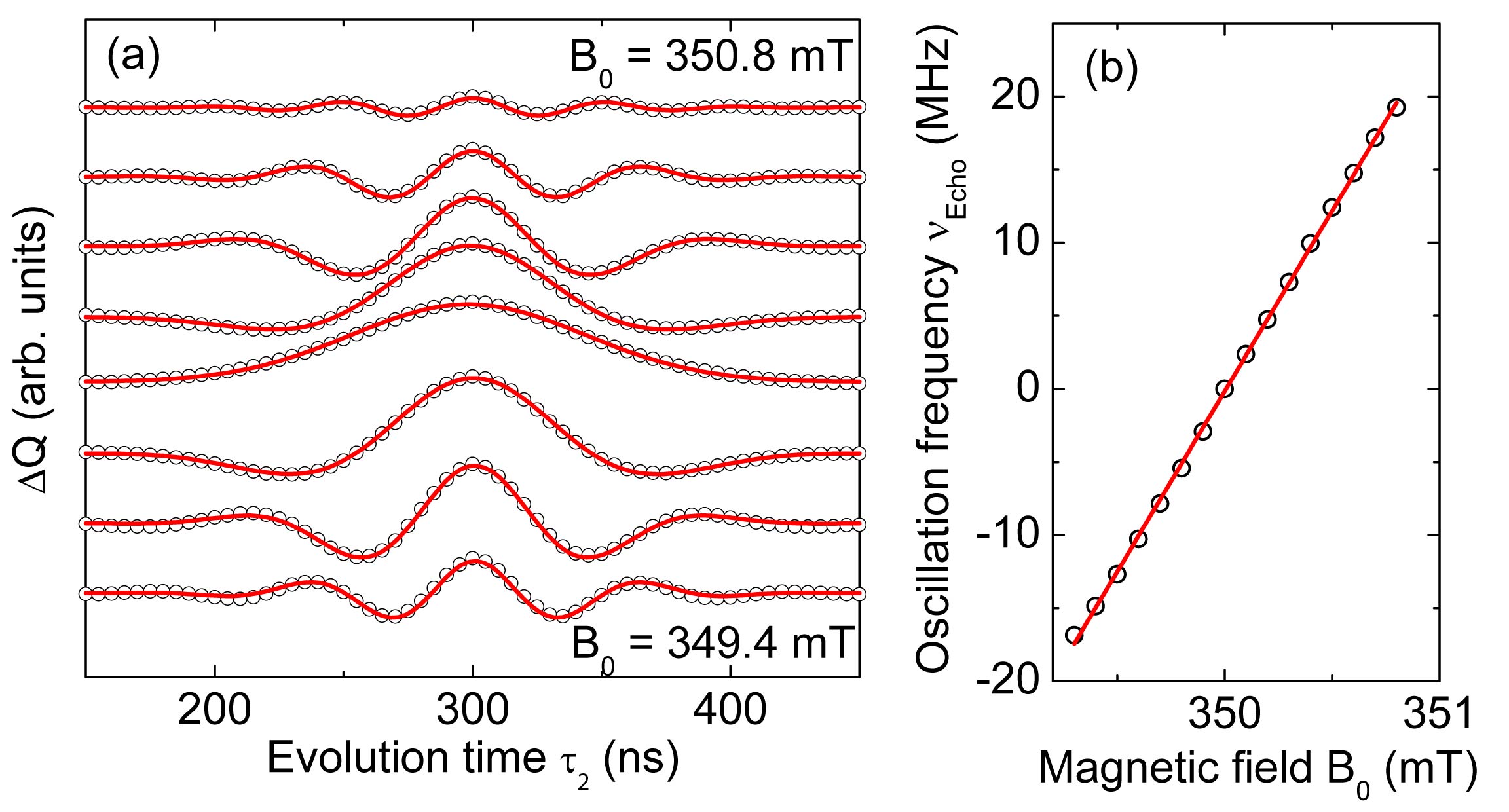} 
\par\end{centering}
\caption{\label{fig:EchotimeOsc}
Oscillations in the evolution time domain.
(a) Cross-sections of $\Delta Q$ along the evolution time axis taken
at different values of $B_{0}$. Red curves are fits based on the
model given by Eq.\,(\ref{eq:DeltaQEcho}). (b) The frequency $\nu_{\mathrm{Echo}}$ of
the damped oscillations as a function of the external magnetic field
$B_{0}$ shows a linear behavior as expected from Eq.\,(\ref{eq:nuEcho}).}
\end{figure}

Similar to the analysis of the EDFID experiment,
various cross-sections of $\Delta Q$ along the evolution time axis
are taken and shown in Fig.\,\ref{fig:EchotimeOsc}(a). Data fitting
based on Eq.\,(\ref{eq:DeltaQEcho}) is performed and illustrated
by the red curves. The oscillation frequency $\nu_{\mathrm{Echo}}$
obtained from the fits is plotted as a function of the external magnetic
field $B_{0}$ {[}Fig.\,\ref{fig:EchotimeOsc}(b){]}, where a linear
dependence of $\nu_{\mathrm{Echo}}$ as a function of $B_{0}$ can
be observed as expected from
\begin{equation}
\nu_{\mathrm{Echo}}=\frac{1}{2\pi}\frac{\bar{\omega}_{1}^{2}}{\sigma_{\omega}^{2}+\bar{\omega}_{1}^{2}}\left(\omega_{0}-\omega_{\mathrm{mw}}\right)=k'\left(B_{0}-B_{\mathrm{res}}\right),\label{eq:nuEcho}\end{equation}
with $k'=g\mu_{\mathrm{B}}/[h(1+4(\sigma_{\omega}/\omega_{1})^{2})]$ analogous to Eq.\,(\ref{eq:nuramseyMF}). The exponential term in Eq.\,(\ref{eq:DeltaQEcho}) describes a Gaussian envelope in the time domain with full width at half maximum (FWHM) 
\begin{equation}
W_{\mathrm{echo}}=2\sqrt{2\ln2}\sqrt{\frac{1}{\sigma_{\omega}^{2}}+\frac{4}{\omega_{1}^{2}}}\,.\label{eq:FWHMEcho}\end{equation} From the fits of the data, an average value of $W_{\mathrm{echo}}=106.9\pm4$\,ns is obtained, corresponding to an expected cwEDMR linewidth of $\Delta B_{\mathrm{PP}}^{\mathrm{Echo}}=0.28\pm0.01$\,mT.
%
This value is consistent with the values $\Delta B_{\mathrm{PP}}^{\mathrm{cwEDMR}}=0.30\pm0.03$\,mT
and $\Delta B_{\mathrm{PP}}^{\mathrm{FID}}=0.26\pm0.02$\,mT obtained
from previous experiments within the accuracy limits. Therefore, the
ED Hahn echo response on the high-field hf($^{31}$P) resonance shows
the same fine structure as the EDFID experiment and can be explained
by the same model as expected from the fact that the echo pulse sequence
consists of two FIDs back to back~\cite{Weil07ESR}. However, the small oscillations
seen in EDFID are not observed in the ED Hahn echo response, since
they are fully defocussed due to the additional central $\pi$-pulse.

\subsection{Spin-spin coupling}

\label{ss:Spin-spin coupling}

So far the experimental results have been discussed in the context
of off-resonance oscillations and dephasing due to inhomogeneous line
broadening. In different previous studies~\cite{Fukui2000,Gliesche2008EffectCoupling},
the possible impact of coupling between the partners of the spin pair
on EDMR experiments has been discussed. In the following, results
of a numerical study of the EDFID experiment discussed above are presented,
focussing on the possibility of EDFID to estimate the coupling strength.
\begin{figure}
\noindent \begin{centering}
\includegraphics[width=8cm]{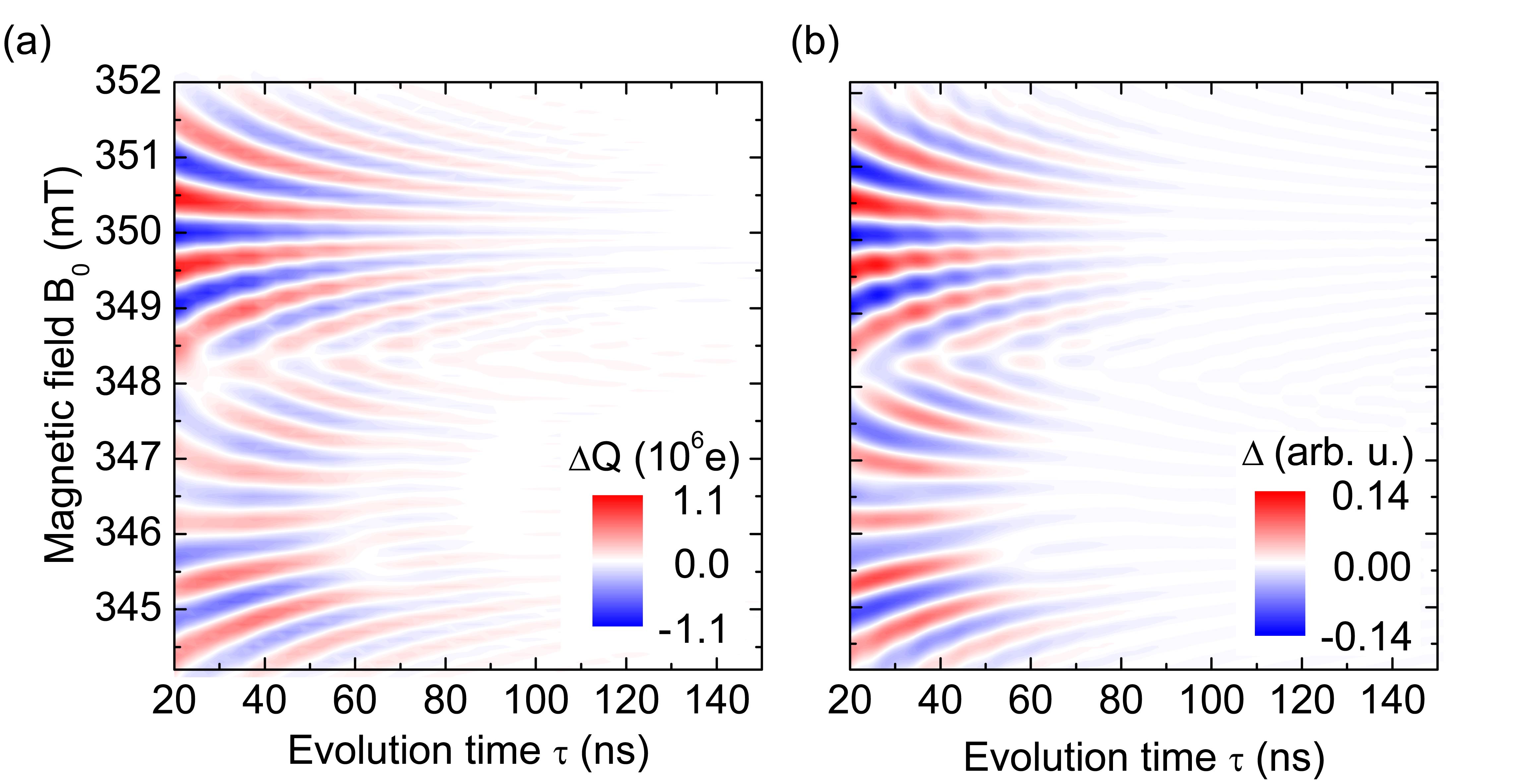} 
\par\end{centering}

\caption{\label{fig:ExpSimu}(a) Experimental results of the EDFID measurement
showing all spectral features. (b) Simulation for $J~=~0$~MHz. In
both simulation and experimental results, the characteristic Ramsey pattern
can be clearly seen at the position of the high-field hf($^{31}$P) resonance,
whereas the patterns at the low-field hf($^{31}$P) and the P$_{\mathrm{b0}}$
resonances are more complicated due to mutual interference. }

\end{figure}

The system is modelled by an ensemble of spin $S=1/2$ pairs described
by the density operator $\hat{\rho}$. The Hamiltonian of an individual
pair is defined as 
\begin{equation}
\hat{\mathcal{H}}=\hat{\mathcal{H}}_{0}+\hat{\mathcal{H}}_{\mathrm{J}}+\hat{\mathcal{H}}_{1}(t)
\label{eq:Hamiltonian}
\end{equation}
with \begin{eqnarray}
\hat{\mathcal{H}}_{0} & = & \frac{1}{2}g_{\mathrm{P}}\mu_{\mathrm{B}}\left(B_{0}\pm B_{\mathrm{HF}}/2+B_{\mathrm{SHF}}\right)\hat{\sigma}_{z}^{\mathrm{P}}\nonumber \\
 &  & +\frac{1}{2}g_{\mathrm{db}}\mu_{\mathrm{B}}\left(B_{0}+B_{\mathrm{\Delta db}}\right)\hat{\sigma}_{z}^{\mathrm{db}}\label{eq:StaticHamilton}\end{eqnarray}
representing the static uncoupled Hamiltonian in the presence of a
constant magnetic field $\mathbf{B}_{0}=B_{0}\mathbf{e}_{z}$ superimposed
with the hyperfine field of $^{31}$P $B_{\mathrm{HF}}=4.2$\,mT
and the superhyperfine field $B_{\mathrm{SHF}}$ at the position of
the donor, where the latter can be considered fixed for timescales
shorter than the precession period of $^{29}$Si nucleus \cite{Lee05Deph}.
$B_{\mathrm{\Delta db}}$ is the local shift of the static magnetic
field at the position of the P$_{\mathrm{b0}}$ center due to effects
such as disorder and superhyperfine interactions. The $\hat{\sigma}_{x,y,z}$
denote the Pauli spin operators. The circularly polarized microwave
of angular frequency $\omega_{\mathrm{mw}}$ and magnitude $B_{1}$
is represented in the rotating frame by 
\begin{equation}
\hat{\mathcal{H}}_{1}=\mu_{\mathrm{B}}B_{1}\left(g_{\mathrm{P}}\hat{\sigma}_{x}^{\mathrm{P}}+g_{\mathrm{db}}\hat{\sigma}_{x}^{\mathrm{db}}\right),
\label{eq:H1}
\end{equation}
which is nonzero during the pulse. Spin-spin interaction is modelled
by an exchange coupling Hamiltonian represented by 
\begin{equation}
\hat{\mathcal{H}}_{\mathrm{J}}=\hbar J\hat{\boldsymbol{\sigma}}^{\mathrm{P}}\cdot\hat{\boldsymbol{\sigma}}^{\mathrm{db}}/4,
\label{eq:Hcoupling}
\end{equation}
with $\hat{\boldsymbol{\sigma}}=(\hat{\sigma}_{x},\hat{\sigma}_{y},\hat{\sigma}_{z})^{T}$.
The dipolar coupling is smaller than 1\,MHz for interspin distances larger than
3\,nm, which is neglected here for simplicity.
The simulation of the spin pair ensemble dynamics is based on the
Liouville equation $\partial_{t}\hat{\rho}=i[\hat{\rho},\hat{\mathcal{H}}]^{-}/\hbar$
in which, in contrast to Eq.\,(5) of Ref.~\cite{Boehme03EDMR}, all
terms related to incoherent processes are dropped since the time constant
of the fastest incoherent process is more than one order of magnitude
larger than the duration of the pulse sequence, as already mentioned
in Section II. The initial steady state of the density operator is
assumed to be given by the pure triplet state $\hat{\rho}^{\mathrm{S}}=\left(|T_{+}\rangle\langle T_{+}|+|T_{-}\rangle\langle T_{-}|\right)/2$
with $|T_{+}\rangle=|\!\!\uparrow^{\mathrm{db}}\uparrow^{\mathrm{P}}\rangle$
and $|T_{-}\rangle=|\!\!\downarrow^{\mathrm{db}}\!\downarrow^{\mathrm{P}}\rangle$
\cite{Boehme03EDMR}. For triplet recombination rates $r_{\mathrm{T}}$
much smaller than the singlet recombination rate $r_{\mathrm{S}}$,
the observable $Q(\tau)$ reflecting the state of the pair ensemble
at the end of the second $\pi/2$ pulse assumes the form $Q(\tau)\propto-\delta(\tau)=-(\delta\rho_{\uparrow^{\mathrm{db}}\downarrow^{\mathrm{P}}}+\delta\rho_{\downarrow^{\mathrm{db}}\uparrow^{\mathrm{P}}})$~\cite{Gliesche2008EffectCoupling},
where $-\delta\rho_{ii}=-(\rho_{ii}(\tau)-\rho_{ii}^{\mathrm{S}})$
denotes the negative difference between the diagonal elements of the
density matrix at the end of the second $\pi/2$-pulse and the initial
steady state. The negative sign expresses the quenching of the photocurrent
due to recombination. Inhomogeneous line broadening is taken into
account by calculating $-\delta(\tau)$ for a single spin pair and
subsequent averaging over Gaussian distributions for both $B_{\mathrm{SHF}}$
and $B_{\Delta\mathrm{db}}$ with experimentally obtained standard
deviations from the pulsed EDFID spectrum shown in Fig.\,\ref{fig:ExpSimu}.
Furthermore, the simulation takes all four combinations of spin pair formation
into account, which arise from the two resonance positions of $^{31}$P, the P$_{\mathrm{b0}}$, and
the P$_{\mathrm{b1}}$.
%
\begin{figure}
\noindent \begin{centering}
\includegraphics[width=8cm]{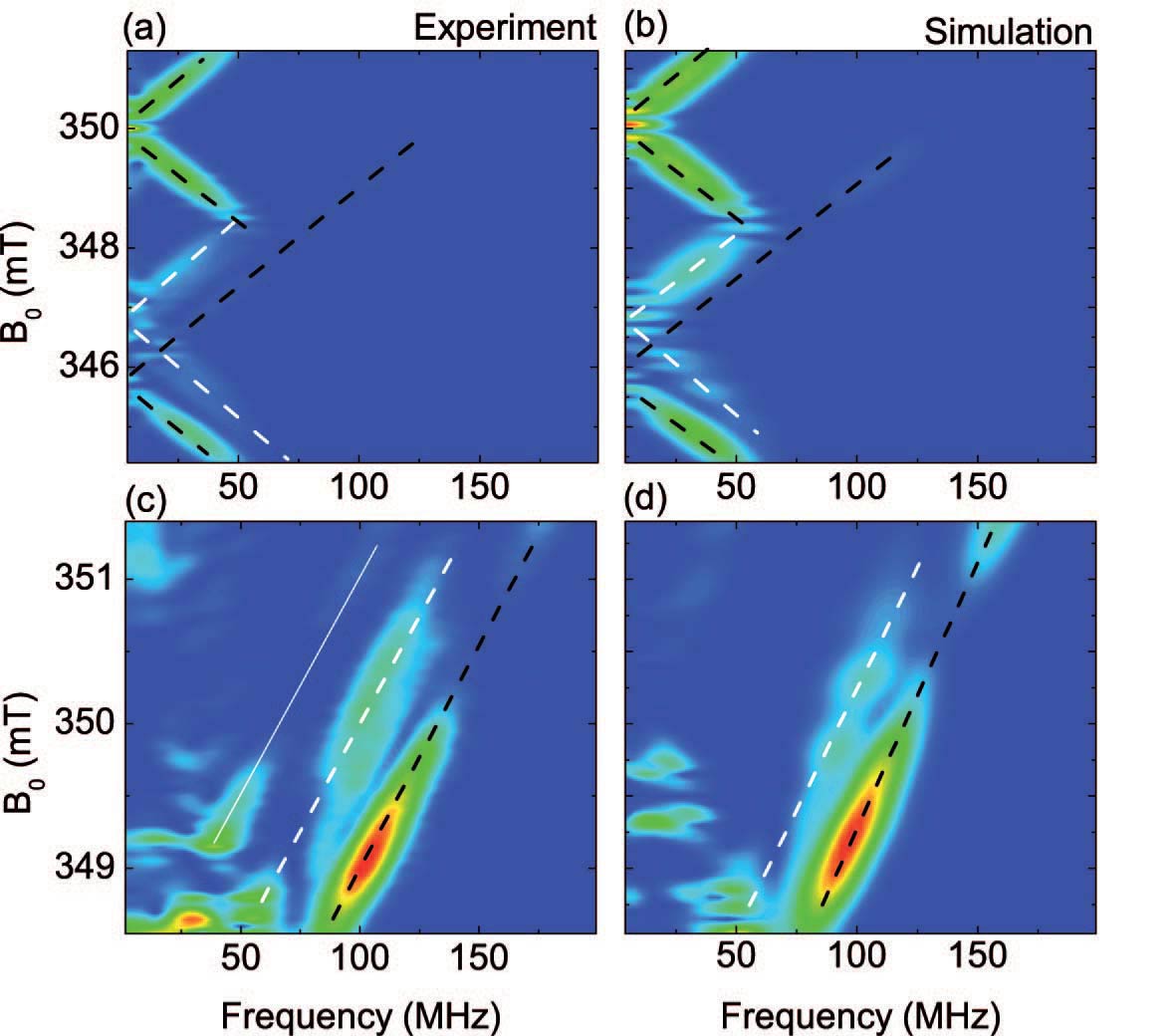} 
\par\end{centering}

\caption{\label{fig:FFT}(a) and (b) Fourier transformation of the EDFID tomography
experiment and simulation shown in Fig.\,\ref{fig:ExpSimu}. The
linear dependence of the oscillation frequency on $B_{0}-B_{\mathrm{res}}$
described by Eq.\,(\ref{eq:nuramseyMF}) is clearly seen in the
frequency domain. The two $^{31}$P resonances and the broader P$_{\mathrm{b0}}$
resonance are marked by black and white dashed lines, respectively.
(c) and (d) For better visibility, details of the FFT-spectrum near
the high-field hf($^{31}$P) resonance are shown after subtracting a background
of the form of Eq.\,(\ref{eq:Ramseyosctime}) from the experimental
and simulated data. Again, the low-field hf($^{31}$P) resonance and the
P$_{\mathrm{b0}}$ resonance are marked by black and white dashed
lines. An additional resonance, indicated by the solid white line,
can be seen in the experimental data. Its spectral position is in accordance
with the small central line observed in cwEDMR (see Fig.\,\ref{fig:cwSpektrum}).}

\end{figure}

Figure\,\ref{fig:ExpSimu}(a) shows the complete experimental results
encompassing all resonances of the EDFID tomography experiment discussed
in Fig.\,\ref{fig:RamseyResult}(a). Figure\,\ref{fig:ExpSimu}(b)
shows the simulation of $-\delta$ for $J=0$ as a function of $B_{0}$
and $\tau$ after subtraction of a constant background obtained from
the value of $-\delta$ for large $\tau$, resulting in the quantity
$\Delta$ which can be compared to $\Delta Q$ in the experiment.
The characteristic patterns of simulation and experiment fit quite well. At the high-field
$^{31}$P resonance, small oscillations superimposed on the Ramsey
oscillation pattern can be seen in the experimental data as well as
in the simulation. These small oscillations are due to the partial
excitation of the low-field $^{31}$P and the P$_{\mathrm{b0}}$ spins
by the microwave pulses on the high-field $^{31}$P resonance. Details
of these patterns are shown in the Fourier transformed data shown
in Fig.\,\ref{fig:FFT} (a) and (b). The linear dependence of the
oscillation frequency on $B_{0}-B_{\mathrm{res}}$ described by Eq.\,(\ref{eq:nuramseyMF})
is clearly visible in the frequency domain. The two $^{31}$P resonances
and the P$_{\mathrm{b0}}$ resonance are marked by black and
white dashed lines, respectively. The P$_{\mathrm{b1}}$ resonance is not resolved due to the spectral overlap with the P$_{\mathrm{b0}}$ and its smaller amplitude. For better visibility, details of
the FFT-spectrum near the high-field hf($^{31}$P) resonance are shown in panels
(c) and (d) after subtracting a background of the form of Eq.\,(\ref{eq:Ramseyosctime})
from the experimental and simulated data. Again, the low-field hf($^{31}$P)
resonance and the P$_{\mathrm{b0}}$ resonance are marked
by black and white dashed lines. An additional resonance, indicated
by the solid white line, can be seen in the experimental data. Its
spectral position is in accordance with the small central line observed
in cwEDMR (see Fig.\,\ref{fig:cwSpektrum}). This line is not taken
into account in the simulation. The intensity of the Fourier amplitude of
the lines as a function of the magnetic field can be described by
an equation of the form $\sin^{2}\left(\frac{\pi}{4}\sqrt{1+x^{2}}\right/(1+x^{2}))$,
analogous to Eq.\,(\ref{eq:Ramseyresult}). In particular, the minima
at frequencies of $\approx65$\,MHz and $\approx130$\,MHz correspond
to rotations of the spins by integer multiples of $2\pi$.

\begin{figure}
\noindent \begin{centering}
\includegraphics[width=8cm]{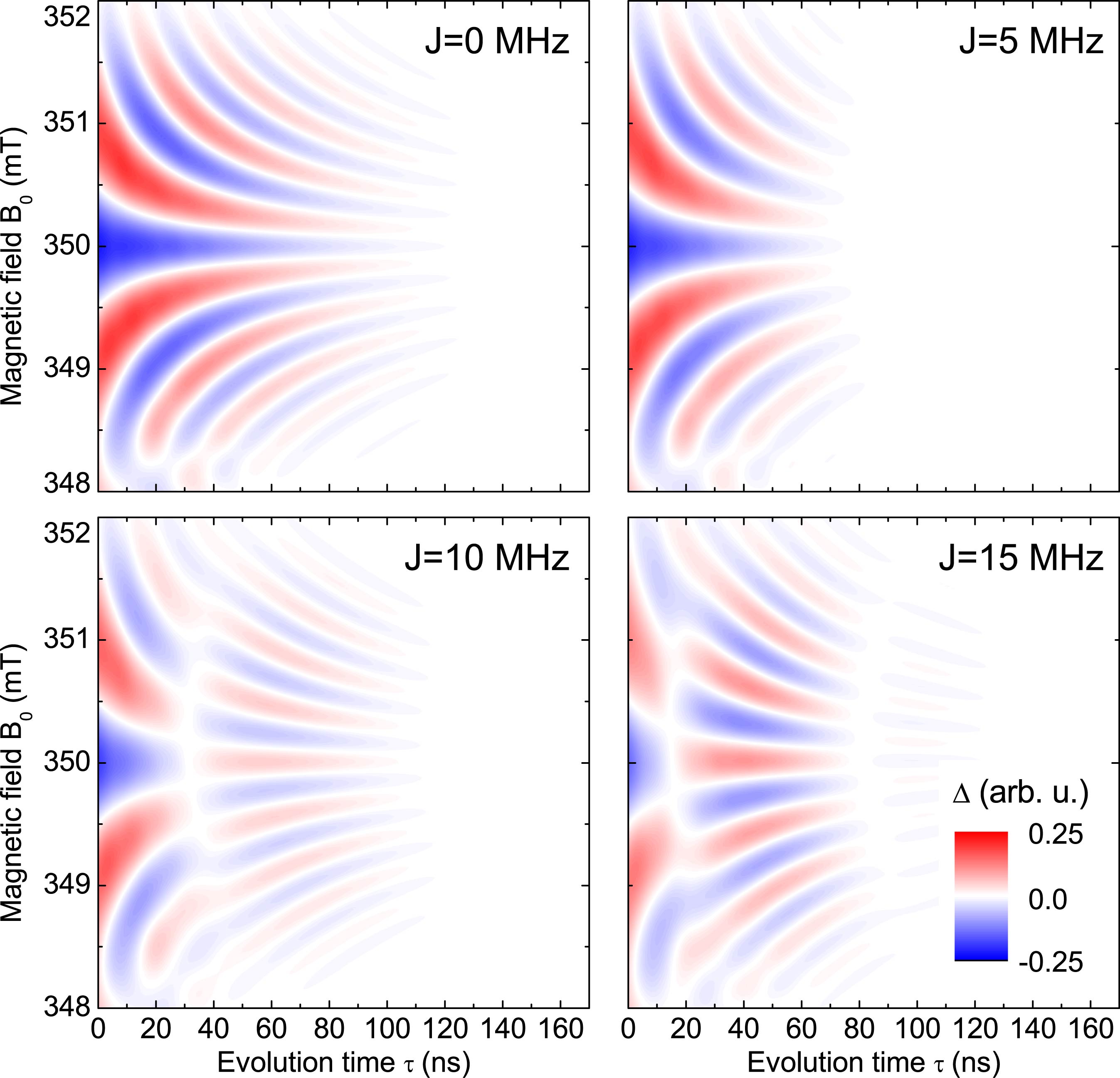} 
\par\end{centering}

\caption{\label{fig:SimulationJ}Simulation of the EDFID experiment at the
high-field hf($^{31}$P) resonance for different exchange coupling parameter.
An oscillation of the signal in the time-domain with the coupling
frequency is expected. This oscillation is masked by the exponential
decay of the signal due to dephasing and therefore can only be resolved
for $J~>~5~$MHz.}

\end{figure}

\begin{figure}
\noindent \begin{centering}
\includegraphics[width=8cm]{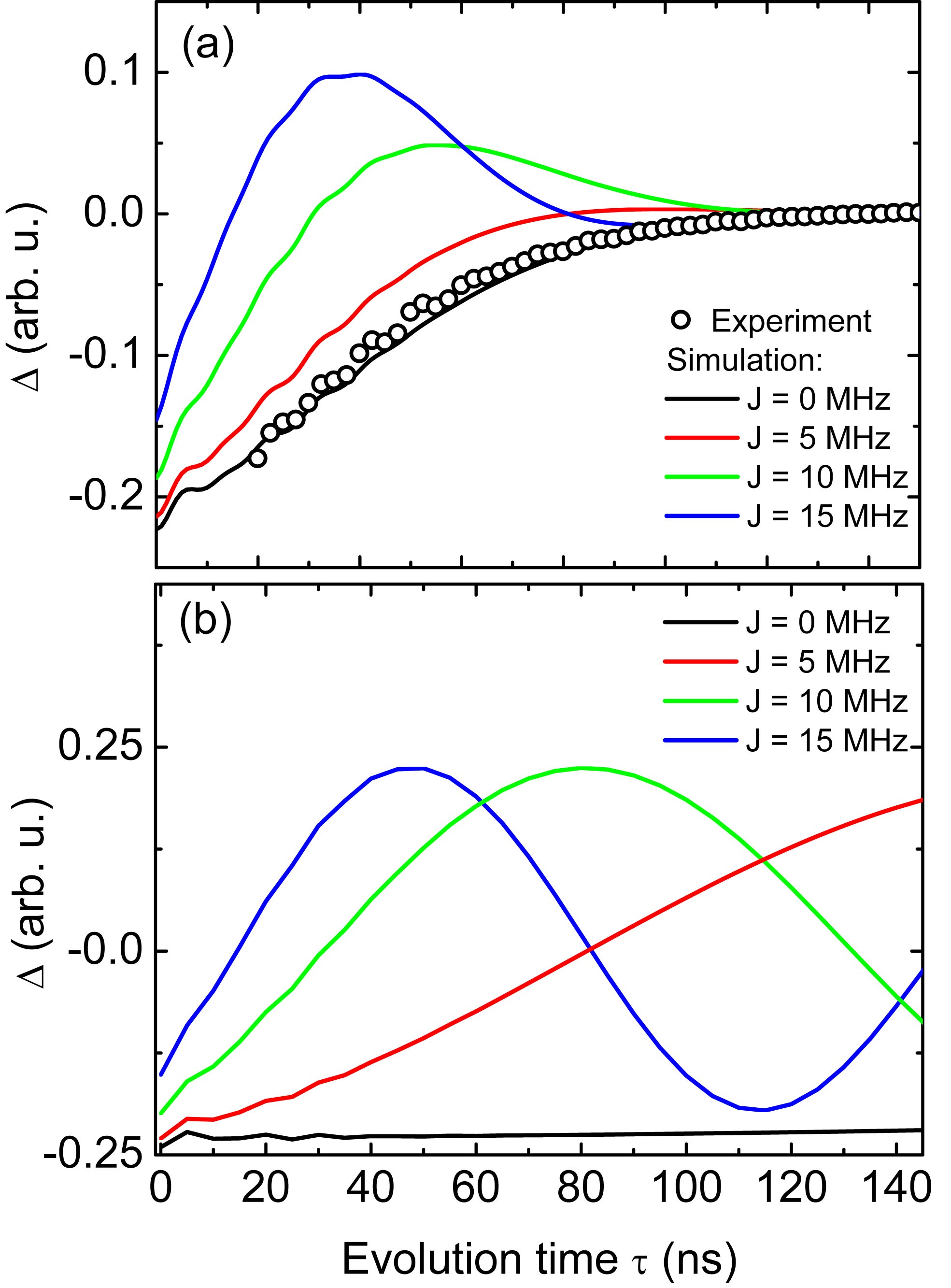} 
\par\end{centering}

\caption{\label{fig:SimuCrosssection} (a) Cross section along the $\tau$
axis at the high-field resonance field of $^{31}$P of the experimental
data (open circles) and the simulated data (solid lines) for different
exchange coupling parameter between $^{31}$P and P$_{\mathrm{b0}}$.
(b) Simulated data like in (a) for an isotopically purified $^{28}$Si
sample. Here, more oscillations caused by the weak coupling can be seen due to
the longer dephasing time.}

\end{figure}

The simulation is further extended to nonzero coupling parameters
with the focus on the clearly observable pattern structure on the
high-field hf($^{31}$P) resonance, which is shown in Fig.\,\ref{fig:SimulationJ}.
Starting from $J=0$, the exchange coupling parameter is increased
in steps of $5$\,MHz resulting in a change of the qualitative behavior of
the characteristic pattern. Whereas for $J=0$ each hyperbola (cf.~Fig.\,\ref{fig:RamseyResult}) either indicates positions of local
maxima or minima, the values of $\Delta$ on each hyperbola oscillate
as a function of $\tau$ for $J\neq0$. This behavior can be clearly
observed on the axis of symmetry at $B_{0}=B_{\mathrm{res}}$, which
is displayed in Fig.\,\ref{fig:SimuCrosssection}(a) for different
values of $J$.%

For vanishing coupling, $\Delta$ relaxes exponentially to the equilibrium.
For larger values of $J$, damped oscillations of $\Delta$ with frequency
$J$ are formed decaying to the equilibrium within the dephasing time
$T_{\mathrm{FID}}$. Compared with the existing experimental data
{[}cf.~Fig.\,\ref{fig:Ramseyfreq}\,(a){]}, the coupling is estimated
$\ll5$\,MHz. Please note, that in these simulations only the high-field
$^{31}$P and the P$_{\mathrm{b0}}$ are taken into account. Therefore, the small oscillations
in Fig.\,\ref{fig:SimuCrosssection} are due to off-resonance excitations
of the P$_{\mathrm{b0}}$ spins.

Clearer insight might be obtained in an experiment using isotopically
purified $^{28}$Si samples as shown in the simulation in Fig.\,\ref{fig:SimuCrosssection}(b)
where the line broadening (although determined by a homogeneous line
width) is qualitatively modelled by a Gaussian distribution with FWHM
line width of $0.023$\,mT corresponding to a concentration of $^{29}$Si nuclei of $\approx$ \% \cite{Abe05Linewidth}. Since $T_{\mathrm{FID}}\sim1/\sigma_{\omega}$
{[}cf. Eq.\,(\ref{eq:Ramseytimeconst}){]} oscillations on the axis
of symmetry could be observed within the dephasing time also for weak $J$. Qualitatively, the frequency of the oscillations increases
with increasing $J$ as evident from Fig.\,\ref{fig:SimuCrosssection}(b).
In the general case, distributions of exchange interaction as well
as dipolar interaction have to be taken into account \cite{Boehme03EDMR},
which would result in an averaging out of the oscillation pattern.
However, previous studies have revealed that only spin-pairs within
a narrow range of intra-pair distances contribute to the observed
signals \cite{Paik2010T1T2}, which should make an estimate of $J$
of contributing spin pairs still possible.

\section{Summary}

To summarize, we have used pulsed EDMR to study the free induction
decay of phosphorus donor spins in silicon. We can resolve oscillations
up to 150~ns limited by dephasing due to superhyperfine interactions
with surrounding $^{29}$Si nuclei. An analytical model is used
to describe the FID of an inhomogeneously broadened line which is
in good agreement with the experimental data. In addition, structures
on two-pulse electron spin echoes have been measured which can be
described by the same analytical model. The results of a numerical
calculation are further presented and compared with the experimental
data to assess the capability of the method to study spin-spin interactions.
From these results, we can give an upper bound for the coupling parameter
of $J~\approx~5$~MHz in the samples studied. 
\begin{acknowledgments}
The work was financially supported by DFG (Grant No. SFB 631, C3) with additional support by the BMBF via EPR-Solar. 
\end{acknowledgments}
\appendix

\section{analytical expression describing the edfid pattern\label{sec:App}}

In this section, Eq.\,(\ref{eq:hyperbolas}) used to describe the
pattern in Fig.\,\ref{fig:RamseyResult}(a) is derived. Neglecting
spin-spin interactions and incoherent processes, the singlet content
$S(\tau)$ is proportional to the flipping probability $P_{\uparrow,\downarrow}$
of a single spin after a $\pi/2$-$\tau$-$\pi/2$ pulse sequence,
which has been investigated in studies related to nuclear magnetic
resonance \cite{Jaynes55matrix,Bloom55Echo}
\begin{eqnarray}
P_{\uparrow,\downarrow}(\tau) & = & 4\sin^{2}\theta\sin^{2}\left(\frac{at_{\mathrm{p}}}{2}\right)\times\left[\cos\left(\frac{\lambda\tau}{2}\right)\cos\left(\frac{at_{\mathrm{p}}}{2}\right)\right.\nonumber \\
 &  & \left.-\cos\theta\sin\left(\frac{\lambda\tau}{2}\right)\sin\left(\frac{at_{\mathrm{p}}}{2}\right)\right]^{2}\label{eq:Ramseyresult}
 \end{eqnarray}
with \[\lambda=\omega_{\mathrm{S}}-\omega_{\mathrm{mw}},\quad a=\sqrt{\lambda^{2}+\omega_{1}^{2}},\quad\sin\theta=\frac{\omega_{1}}{a},\]
where $\omega_{\mathrm{S}}$ is the Larmor frequency of the $^{31}$P
donor electron and $\omega_{\mathrm{mw}}$ the microwave frequency.
$t_{\mathrm{p}}=\pi/(2\omega_{1})$ and $\tau$ denote the length
of the $\pi/2$ pulse and the free evolution time, respectively. For
inhomogenously broadened lines, the observable $S^{\mathrm{av}}(\tau)$
is obtained by averaging $S(\tau)\propto P_{\uparrow,\downarrow}(\tau)$
over the Larmor frequency distribution \cite{Boehme03EDMR} 
\begin{equation}
S^{\mathrm{av}}(\tau)\propto\int\limits _{-\infty}^{\infty}\Phi(\omega_{\mathrm{S}})P_{\uparrow,\downarrow}(\tau,\omega_{\mathrm{S}})\mathrm{d}\omega_{\mathrm{S}}.\label{eq:Savr}
\end{equation}
For distributions $\Phi(\omega_{\mathrm{S}})$ with a maximum at
the center frequency $\omega_{0}$, the dominant term of $S^{\mathrm{av}}(\tau)$
for $\omega_{\mathrm{mw}}$ close to $\omega_{0}$ is given by 
\begin{equation}
S^{\mathrm{av}}(\tau)\propto\int\limits _{-\infty}^{\infty}\Phi(\omega_{\mathrm{S}})\frac{\sin^{2}\left(\frac{\pi}{2}\sqrt{1+x^{2}}\right)}{1+x^{2}}\frac{1+\cos\left(\lambda\tau\right)}{2}\mathrm{d}\omega_{\mathrm{S}},\label{eq:Seff}\end{equation}
 with $x=\lambda/\omega_{1}$, neglecting a term of $\mathcal{O}(|\omega_{\mathrm{mw}}-\omega_{S}|^{2})$.
To obtain an analytical expression, Eq.\,(\ref{eq:Seff}) can be further
simplified by the approximation
\begin{equation}
\frac{\sin^{2}\left(\frac{\pi}{2}\sqrt{1+x^{2}}\right)}{1+x^{2}}\approx\exp(-x^{2})\label{eq:approx}
\end{equation}
since both functions share the same leading orders in the Taylor
expansion, tolerating a deviation of 6\% in the integrated area within
the interval defined by the zero-crossings of $\sin^{2}(...)/(1+x^{2})$
in Eq.\,(\ref{eq:Seff}). Modelling the Larmor frequency distribution
by a Gaussian 
\begin{equation}
\Phi(\omega_{\mathrm{S}})=\frac{1}{\sqrt{2\pi}\sigma_{\omega}}\exp\left[-\frac{1}{2}\left(\frac{\omega_{\mathrm{S}}-\omega_{0}}{\sigma_{\omega}}\right)^{2}\right]\label{eq:gaussian}
\end{equation}
with standard deviation $\sigma_{\omega}$ and center $\omega_{0}$,
the average singlet content is given by 
\begin{eqnarray}
S^{\mathrm{av}} & \propto & \exp\left[-\frac{1}{2}\frac{\Delta\omega^{2}}{\sigma_{\omega}^{2}+\tilde{\omega}_{1}^{2}}\right]\left\{ 1+\exp\left[-\frac{1}{2}\frac{\sigma_{\omega}^{2}\tilde{\omega}_{1}^{2}}{\sigma_{\omega}^{2}+\tilde{\omega}_{1}^{2}}\tau^{2}\right]\right.\nonumber \\
 &  & \left.\times\cos\left[\frac{\tilde{\omega}_{1}^{2}}{\sigma_{\omega}^{2}+\tilde{\omega}_{1}^{2}}\Delta\omega\tau\right]\right\} \label{eq:ramseysignal2}
\end{eqnarray}
with $\tilde{\omega}_{1}=\omega_{1}/\sqrt{2}$ and $\Delta\omega=\omega_{0}-\omega_{\mathrm{mw}}$.
Since $Q\propto-S^{\mathrm{av}}$, $\Delta Q$ is proportional to $-\left[S^{\mathrm{av}}(\tau)-S^{\mathrm{av}}(\tau\to\infty)\right]$ as
the constant background given by $S^{\mathrm{av}}(\tau\to\infty)$ is identical for the signals obtained for both phases (+x and -x) of the last $\pi/2$ pulse and thus subtracted by the data evaluation procedure described in Sec.~\ref{ss:ExpDetails}. This results in
\begin{equation}
\Delta Q\propto-\exp\left[-\frac{1}{2}\frac{\sigma_{\omega}^{2}\tilde{\omega}_{1}^{2}}{\sigma_{\omega}^{2}+\tilde{\omega}_{1}^{2}}\tau^{2}\right]\cos\left[\frac{\tilde{\omega}_{1}^{2}}{\sigma_{\omega}^{2}+\tilde{\omega}_{1}^{2}}\Delta\omega\tau\right]\label{eq:ramseysig2}
\end{equation}
with local extrema approximately determined by values of $B_{0}$
and $\tau$ for which the cosine term in Eq.\,(\ref{eq:ramseysig2})
is equal to $\pm1$, i.e.
\begin{equation}
B_{0}-B_{\mathrm{res}}=\frac{n\pi\hbar\left(1+2(\sigma_{\omega}/\omega_{1})^{2}\right)}{g\mu_{B}}\frac{1}{\tau},\qquad n\in\mathbb{Z}.\label{eq:hyperbolas2}
\end{equation}
This term represents hyperbolas in the $B_{0}$-$\tau$ plane shown
in Fig.\,\ref{fig:RamseyResult}.


\begin{thebibliography}{41}
\bibitem[{fac()}]{factorsqrt2}
\bibinfo{howpublished}{The factor of $\sqrt{2}$ between
  $\tilde{\omega}_{1}$~[Eq.\,(\ref{eq:ramseysig})] and
  $\bar{\omega}_{1}$~[Eq.\,(\ref{eq:DeltaQEcho})] can be explained by the
  different excitation profiles of the corresponding pulse sequences. Since the
  spin-echo pulse sequence consists of two EDFID pulse sequences, the effective
  excitation profile of the EDFID pulse sequence [Eq.\,(\ref{eq:approx})] has
  to be squared to described the spin echo. Hence, its excitation bandwidth is
  reduced by a factor of $\sqrt{2}$ since
  $[\exp(-x^2)]^2=\exp[-(x/(1/\sqrt{2})^2]$}.
\end{thebibliography}

\end{document}